\documentclass{kluwer}    
\usepackage{epsfig}

\newdisplay{guess}{Conjecture}

\begin{document}                                                                                   
\begin{article}
\begin{opening}         
\title{Spectral Indices of  Stars at Super-solar Regime\thanks{Based on 
observations collected at the INAOE ``G. Haro'' Observatory, Cananea (Mexico).}} 

\author{Alberto \surname{Buzzoni}}  
\institute{Telescopio Nazionale Galileo, A.P. 565, 38700 Santa Cruz de La Palma (Tf), Spain,\\
and Osservatorio Astronomico di Brera, Milano Italy}

\author{Miguel \surname{Chavez}}  
\institute{Instituto Nacional de Astrof\'\i sica, Optica y Electr\'onica, A.P. 51 y 216, 72000 Puebla, Mexico}

\author{Maria Lucia \surname{Malagnini}}  
\institute{Dipartimento di Astronomia, Universit\`a degli Studi di Trieste, Via G.B. Tiepolo 11, 34131 Trieste, Italy}

\author{Carlo \surname{Morossi}}  
\institute{Osservatorio Astronomico di Trieste, Via G.B. Tiepolo 11, 34131 Trieste, Italy}

\runningauthor{Buzzoni et al.}
\runningtitle{Spectral Indices of Stars at Super-solar Regime} 


\begin{abstract}
We derived Lick narrow-band indices for  139 candidate super metal-rich stars of 
different luminosity class for a sample of which fundamental atmosphere parameters 
have been obtained
from a grid of synthetic reference spectra by Malagnini et al.\ (2000). 
Indices include Iron Fe50, Fe52, Fe53, and Magnesium Mgb and Mg$_2$ features.
By comparing models and observations, no evidence is found for non-standard
Mg vs.\ Fe relative abundance (i.e.\ [Mg/Fe]~=~0, on the average, for our sample).
Both the Worthey et al. (1994) and Buzzoni et al. (1992, 1994) fitting functions
are found to suitably match the data, and can therefore confidently be extended 
for population synthesis applications also to super-solar metallicity regimes.

\end{abstract}

\end{opening}

\section{The Databases}  

As a useful tool to investigate the distinctive properties of stellar aggregates, 
the fitting-function technique (Buzzoni 1995; Worthey et al.\ 1995)
has been extensively used in population synthesis models to reproduce integrated 
spectroscopic indices, even in super metal-rich (SMR) systems such as elliptical galaxies 
or bulges of spirals.

Since these relations mainly derive from a fit of the local stellar population,
with $Z \sim Z_\odot$, it is important to confidently assess their validity 
also at super-solar metallicity regimes. For this reason, we derived 
Lick narrow-band indices from spectroscopic observations of a sample of 139 SMR stars 
with $[Fe/H] \geq +0.1$, comparing with a suitable theoretical database 
of synthetic reference spectra. A summary of the main characteristics of the databases 
is given in Table~1 (see also Chavez et al.\ 1997, and Malagnini et al.\ 2000 for 
further details).
 
From the observed spectra, Iron Fe50, Fe52, and Fe53, as well as Magnesium Mgb and 
Mg$_2$ indices have been computed according to Worthey et al.\ (1994, hereafter W94).
Calibration to the standard system has been done by using the subset of 49 stars 
in common with the Lick original database, and/or with Buzzoni et al.\ (1992, 1994; 
B92 and B94, respectively).

\begin{table} %
\caption[]{The Databases.}
\begin{tabular}{lcl}                                        
\hline
Observations & ~~~~~~~~~~~&Synthetic Spectra \\
\hline
139 Pop I Stars &  & Chavez et al. 1997 collection  \\
300 spectra     &  & 693 spectra       \\
$4600 \div 5500$~\AA\  &   & $4850 \div 5400$~\AA\  \\
$R=\Delta \lambda/\lambda=2000$   &  & $R=\Delta \lambda/\lambda=250\,000$    \\
$[Fe/H] \ge 0.1$  & & $[M/H] = -1.0~ \to +0.5$  \\
M $\to$ F spectral type & & $4000\ K < T_{eff} < 8000\ K$ \\
I $\to$ V MK luminosity class  &     & log g = 1.0 $\to$ 5.0\\ 
\hline
\end{tabular}
\end{table}

\section{Index residuals and atmosphere parameters} 

The (O--C) index residuals obtained by comparing observations with the corresponding
fitting spectra from the theoretical database have been analyzed  vs.\ atmosphere 
fiducial parameters, according to Mala-gnini et al.\ (2000).
Figure~1 shows the residual distribution for Mg$_2$ and the combined Iron index 
$<{\rm Fe}> = (Fe52+Fe53)/2$ vs.\ [Fe/H]. Solid dots mark the ``fair'' 
subsample of 73 stars with more confident fundamental parameters (cf.\ Malagnini et al.\ 2000).

The four points with the largest (negative) residuals correspond to the coolest stars in 
our sample ($T_{\rm eff} \leq 4000~K$), and reflect the partial inadequacy of the theoretical
database at this temperature range. As far as the remaining sample of 135 stars is considered,
however, the residual distribution for all indices is consistent with a zero-average hypothesis. 
Quite remarkably, we also verified the lack of any negative correlation between 
Mg and Fe (O--C) residuals. This indicates that a non-solar partition for the [Mg/Fe] relative
abundance can confidently be ruled out for our stars.

\begin{figure}
\centerline{ 
\epsfig{figure=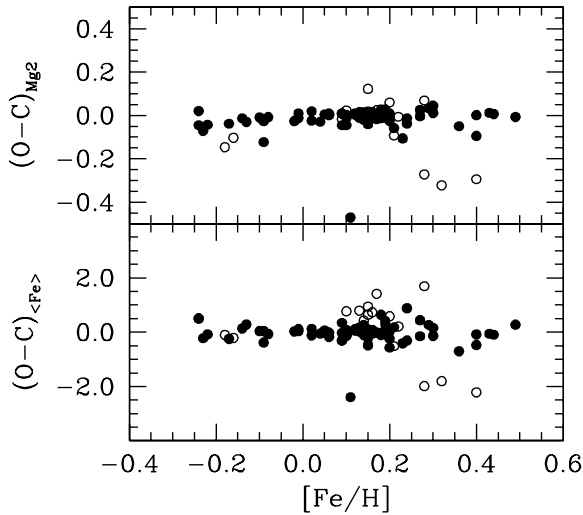,width=0.7 \hsize}
}
\caption{(O--C) index residuals vs.\ [Fe/H]. The 73 stars in the ``fair'' subsample, with more 
confident atmosphere parameters, are singled out (``$\bullet$'' markers).
Excluding the four outliers (the coolest stars in our sample with 
$T_{\rm eff} \leq 4000~K$) (O--C) residuals are consistent with a zero average.}
\end{figure}

\section{Fitting Functions}

Analytical fitting functions, giving the Fe and Mg index strength
vs.\ stellar atmosphere parameters across the H-R diagram, have been provided by 
B92, B94, and W94.
To check self-consistency of these functions also in the SMR regime we studied their
index residuals with respect to our observations.
Figure~2 summarizes the results for Mg$_2$ and $<{\rm Fe}>$. 

The (O--C) distribution in both panels is consistent with a zero average 
residual confirming that the two sets of equations properly account for 
high-metallicity stars, at least in the range of the atmosphere parameters
sampled by our stars.
The W94 fitting function is slightly more accurate than the B92 one
(points spread for the ``fair'' sample is $\sigma({\rm Mg}_2) 
= \pm 0.047$~mag vs. $\pm 0.052$~mag for the latter 
case), but at cost of a more elaborated multi-branch analytical fit.

A similar behaviour is found for the $<{\rm Fe}>$ index,
with $\sigma(<{\rm Fe}>) = \pm 0.61$~\AA\ and $\pm 0.34$~\AA\ with respect 
to B94 and W94, respectively. We verified that the skewed positive
residuals with respect to B94, evident from the figure, mostly 
come from the warmer stars in the sample ($T_{eff} \gtrsim 6700$~K) for which 
the B94 output predicts a vanishing Fe52 index.

\begin{figure}
\epsfig{figure=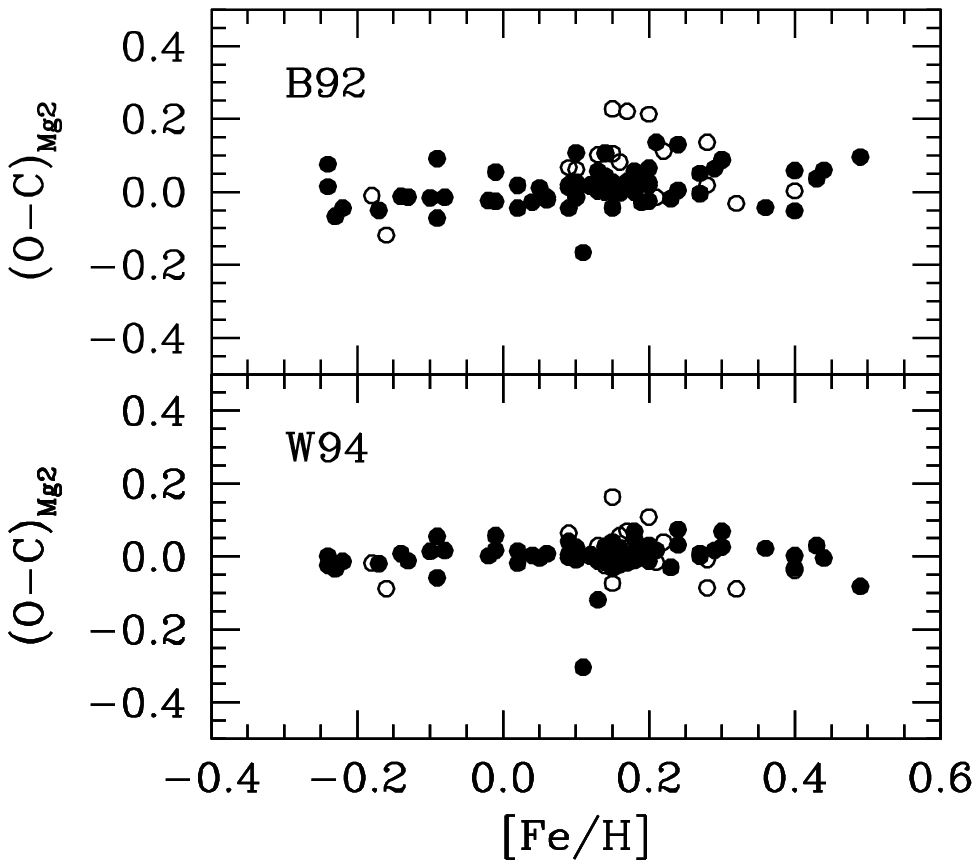,width=0.5 \hsize}
\epsfig{figure=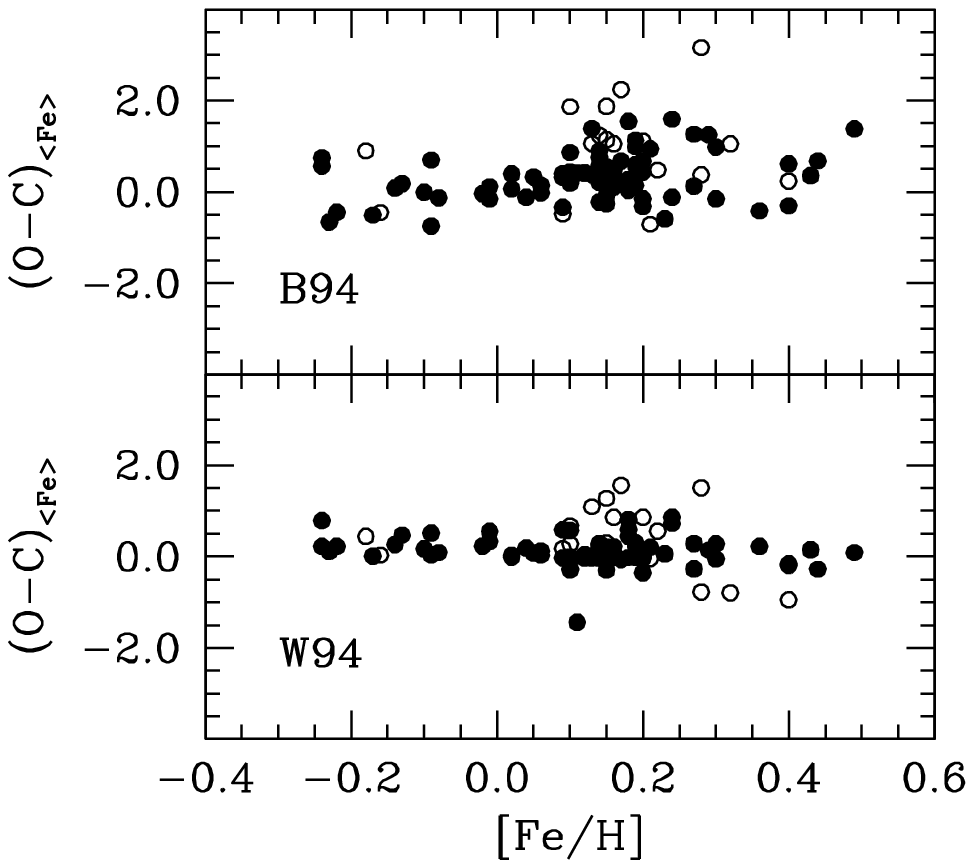,width=0.5 \hsize}
\caption{The Mg$_2$ and $<{\rm Fe}>$ index residuals (in the sense [Observed -- Computed])
with respect to the Buzzoni et al.\ (1994, 1994) and Worthey et al.\ (1994) 
fitting functions are displayed vs.\ metallicity. Like in Fig.~1, solid dots mark
stars in the ``fair'' subsample. Both (O--C) distributions are consistent with a zero average.}
\end{figure}

\acknowledgements
This work received partial financial support from the Italian MURST via COFIN '98/'00, and 
60\% grants, and from the Mexican CONACyT via grant 28506-E.

\end{article}
\end{document}